\documentclass[aps,prl,reprint,groupedaddress]{revtex4-1}

\usepackage{graphicx}
\usepackage{subcaption}
\usepackage{amssymb}
\usepackage{amsmath}
\usepackage{bbold}

\begin{document}
\title{SAW interdigitated transducers as topological mechanical metamaterial}
\author{S. McHugh}
\email[]{smchugh@resonant.com}
\affiliation{Resonant Inc., Santa Barbara, California 93117 USA}
\date{\today}

\begin{abstract}
A lattice model is developed to describe the mechanical displacement of and current through each electrode of a surface acoustic wave (SAW) interdigitated tranducer (IDT).  Each electrode of an IDT is treated as a mass connected mechanically to its neighbors with a spring and electrically with a capacitor.  Simulations for the electrical admittance of a typical SAW IDT are performed and compared with the results of an accurate finite element method simulation.  The utility of this lattice model is demonstrated by simulating the admittance of an IDT structure known as a hiccup resonator, which has a mode in the center of the band gap.  It is shown here that this mode is a topologically protected edge state described by the 1D Su-Schrieffer-Heeger (SSH) model. Hiccup resonators have been used in commercial products for decades, and as such it may considered the first mass-produced topological mechanical metamaterial.
\end{abstract}


\maketitle

The phonon band structure of a natural material is controlled by the material properties.  Acoustic metamaterials have fabricated structures larger than the material unit cell and are designed to modify the natural elastic wave band structure \cite{Cummer}. Although the definitions are somewhat loose, mechanical metamaterials may be thought of as an abstraction of acoustic metamaterials, where the unit cells are composed of discrete elements obeying classical equations of motion.  Assembled in lattices, there exist collective exitations described by dispersion relations and band structure \cite{Surjadi}.  Some mechanical metamaterials allow for the construction of classical analogs to topological electronic band structures seen in 2D materials like graphene.  For instance, the authors in Refs. \cite{Wang, Nash} construct a lattice of coupled gyroscopes engineered to mimic the physics of Quantum Hall sytems.  The gyroscopes break time-reversal symmetry and produce the chiral edge states of Quantum Hall systems as collective mechanical motion of the gyroscopes propagating around the perimeter of the lattice.  As another example, a many-body Hamiltonian for the Quantum Spin-Hall effect\cite{KaneMele} was mapped from a many-body Hamiltonian to a system of equations for classical, coupled harmonic oscillators obeying Newton's laws\cite{SusstrunkHuber}. The topologically protected helical edge states were observed as a collective excitations of pendula.  Similar approaches have been explored by a number of different groups, \cite{KaneLubensky,Pal, Rosa, Kariyado,Salerno,Socolar}.  References \cite{SusstrunkHuber2,BarlasProdan} summarize the various approaches to these linear mechanical metamaterials.  

\begin{figure}[h!]
    \includegraphics[trim=0 0 0 0,clip,width=0.5\textwidth]{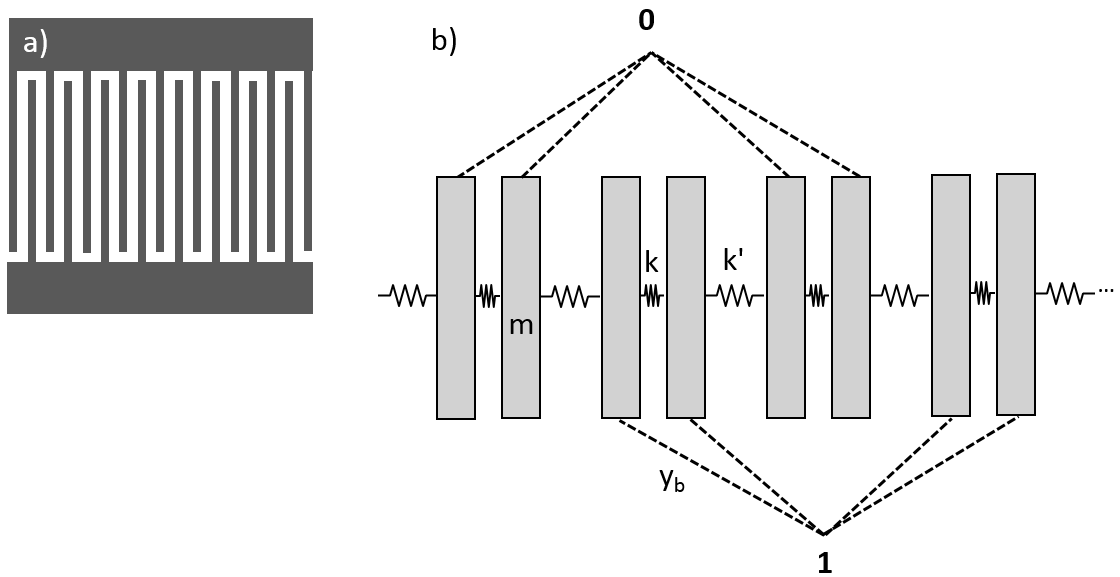}
  \caption{(\textit{a}) Top-view schematic of typical IDT with every other electrode electrically connected via a busbar.  For IDTs with sufficiently long electrodes, an approximately 1D lattice is formed. If the electrode dimensions and gaps between them do not vary along the length of the IDT, it is referred to as a synchronous resonator. (\textit{b}) Illustration of lattice model for synchronous IDT.  Each physical electrode is modeled as two point masses, $m$, mechanically coupled together with spring $k$ and to its neighbor with a spring, $k'$.  Schematically, the magnitudes of $k$ and $k'$ are inversely proportional to their lengths.  Neighboring electrodes are coupled electrically with capacitance $C$, and electromechanically with $\alpha$. The busbar is modeled by connecting every other mass pair to a common electrical node.  Each mass has a mechanical, $u$ and electrical, $I$ degree of freedom.}
\label{schematic_synchronous}
\end{figure}

An IDT may be considered an acoustic metamaterial.  Typically arranged on the surface of a piezoelectric material, as in Figure \ref{schematic_synchronous}a, IDTs form a SAW resonator.  It is well-known that surface waves progating through an IDT exihibit dispersion and an acoustic band gap defined by the periodicity and composition of the electrodes on the surface\cite{Morgan,Hashimoto}.  The purpose of this paper is to show that an IDT may also be considered a mechanical metamaterial.  The focus here is on the mechanical displacements of and currents through the electrodes of the IDTs themselves, rather than the surface waves generated by the IDT.  A simple lattice model is presented for an IDT where each electrode is treated as a discrete mass mechanically and electrically coupled to its neighbors. Although there are excellent methods for simulating the electro-mechanics of an IDT, e.g., equivalent circuits \cite{Butterworth, vanDyke, Thorvaldsson} and coupling-of-modes methods \cite{HausCOM, HausWrightCOM}, using the simple lattice model introduced here makes the connection to existing mechanical metamaterial work clear. The hallmarks of a mechanical metamaterial are present; collective mechanical excitations of the electrodes obey a dispersion relation, a band gap is present, and the character of both can be easily engineered.  A hiccup resonator, which is essentially an IDT with a missing electrode near the center\cite{Wright}, makes for an interesting demonstration. It is characterized by an additional resonance in the acoustic band gap. As shown below, the missing electrode creates a domain wall or edge, which allows for a direct realization of a topologically protected edge state of the 1D SSH model \cite{SSH, KaneLubensky,SusstrunkHuber2}.  The results are confirmed with simulations produced with a commercial FEM software package tailored to produce highly accurate simulations for IDTs \cite{KoskelaLayers}.  

In general, piezoelectric IDTs are an attractive platform to study a variety of mechanical metamaterials.  The large electro-mechanical coupling of common piezoelectric crystals such as LiNbO$_3$ and LiTaO$_3$ makes the ultrasonic mechanical displacements easily generated and sensed with electronic instruments.  The also have high quality factors above 1 GHz at room temperatures and excellent linearity \cite{Hashimoto, Morgan}. It is for these reasons they are used in nearly every mobile phone. Hiccup resonators in particular, have been widely used in voltage control oscillators and narrow band notch filters for over three decades \cite{Wright,WrightReview, Gulyaev, PlesskyHiccup}.  In retrospect, hiccup resonators may rightly be considered the first mass-produced, topological mechanical metamaterial.

The schematic in Fig. \ref{schematic_synchronous}a shows the top view of a ``synchronous'' IDT or resonator, where the electrodes and the gaps between them are all identical and every-other electrode is electrically connected via a ``busbar.''  For sufficiently long electrodes, an approximately 1D lattice is formed.  The essential elements to simulate a synchronous IDT are shown in Fig. \ref{schematic_synchronous}b.  Each electrode is modeled as two point masses, $m$, coupled with spring, $k$, and coupled to the masses of neighboring electrodes with spring, $k'$.  Each mass is also damped with dissipation constant, $\tau$.  The masses form a system of coupled, damped harmonic oscillators.  In addition, there is a capacitance, $C$, formed across the gap between electrodes.  For this synchronous IDT, every other electrode is electrically coupled to a common node with an electrical admittance, $y_b$.  Finally, there is an electro-mechanical coupling constant, $\alpha$ formed between neighboring electrodes.  Each $m$ has a mechanical and electrical degree of freedom, $u$ and $I$, respectively.  The parameters of the lattice model have the intuitive relationship with the real dimensions of the IDT, e.g., $m$ is proportional to the thickness and width of the electrode, and $k', \alpha,$ and $C$ vary inversely with respect to the gap between neighbors.  

The following equations are motivated and derived in the appendix for an IDT with $N/2$ electrodes,
\begin{eqnarray}
i\omega C\mathbf{v} + Y_{bus}\mathbf{v}=\mathbf{I} + i\omega\mathbf{\alpha u}\\
(k+i\omega\tau-m\omega^2)\bf {u}= -\alpha \mathbf{v},
\end{eqnarray}
where $\bf{u}$ = $(u_0, u_1,...u_{N-1})$ and $\bf{I}$ = $(I_0, I_1,...I_{N-1})$ are the degrees of freedom for each mass. $\bf{v}$ = $(v_0, v_1,...v_{N-1})$ are the driving voltages. $C, \alpha, k, \tau$, and $m$ are now considered $N\times N$ matrices.  $Y_{bus}$ is the resistive electrical admittance matrix connecting the electrodes, i.e., busbar. 

Eq. 1 is in the form of Ohm's law with a source proportional to the mechanical displacement. Eq. 2  describes a set of coupled harmonic oscillators with a source proportional to the voltage.  Numerically solving these equations for an arbitrary IDT is straightforward.  The most useful quantity to calculate is the electrical admittance from nodes 0 to 1, $y$, calculated from the matrix, 
 \begin{equation}
 Y = Y_{bus}+i\omega C +i\omega\alpha\left(k+i\omega\tau - m\omega^2\right)^{-1}\alpha,
 \end{equation}  
 which is derived in the appendix.  The three matrices in the parentheses describe the undriven mechanical system, Eq. 2. It is clear from the last term that $y$ is sensitive to the mechanical eigenmodes.  In general, if the dissipation is low, peaks in the conductance, $y_{real}$ will occur near an eigenfrequency.  Note, as demonstrated below, the converse is not true.

\begin{figure}[h!]
	\begin{subfigure}[t]{0.5\textwidth}
		\includegraphics[width=1\textwidth]{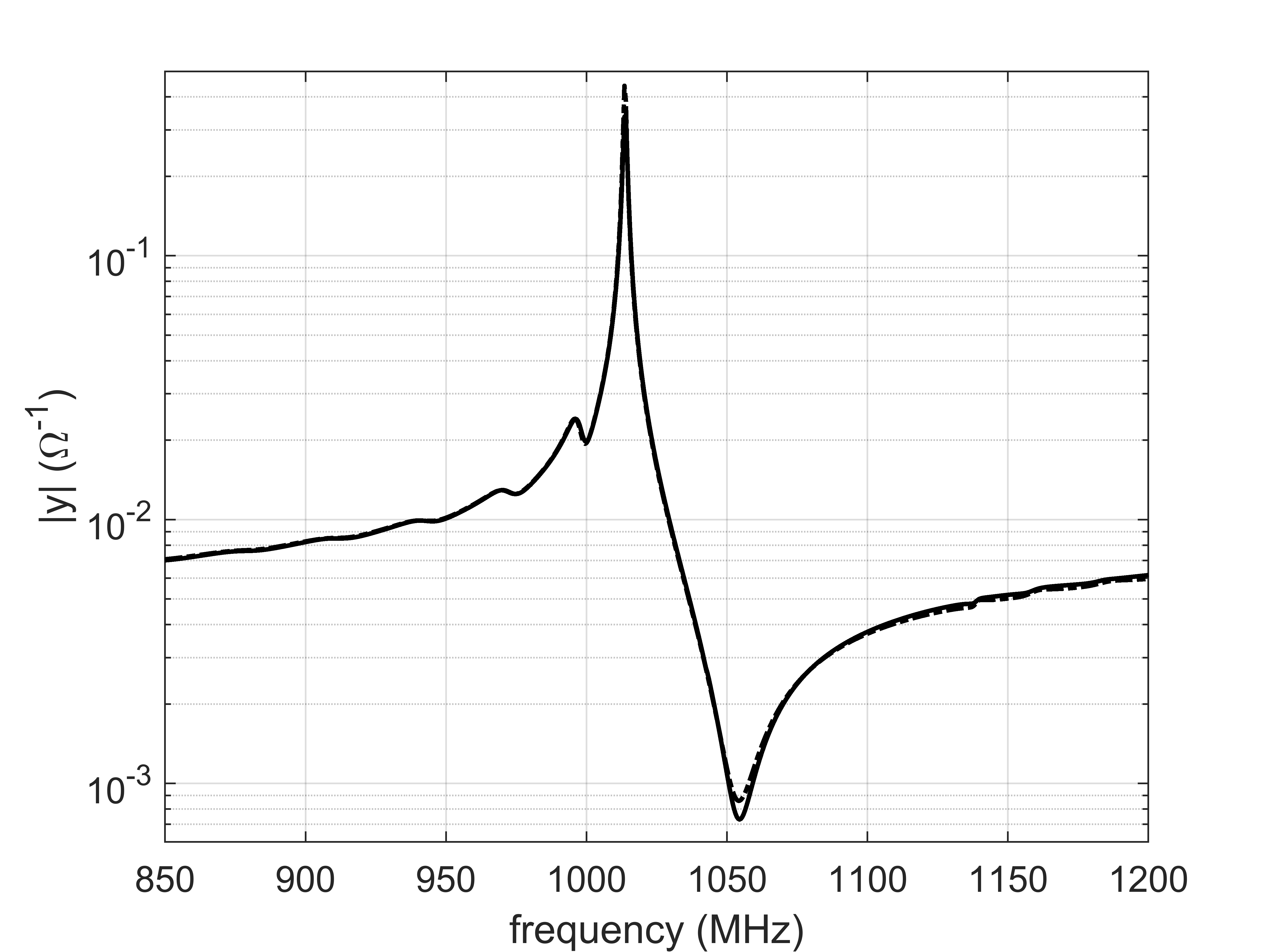}
	\end{subfigure}
	\begin{subfigure}[t]{0.5\textwidth}
		\includegraphics[width=1\textwidth]{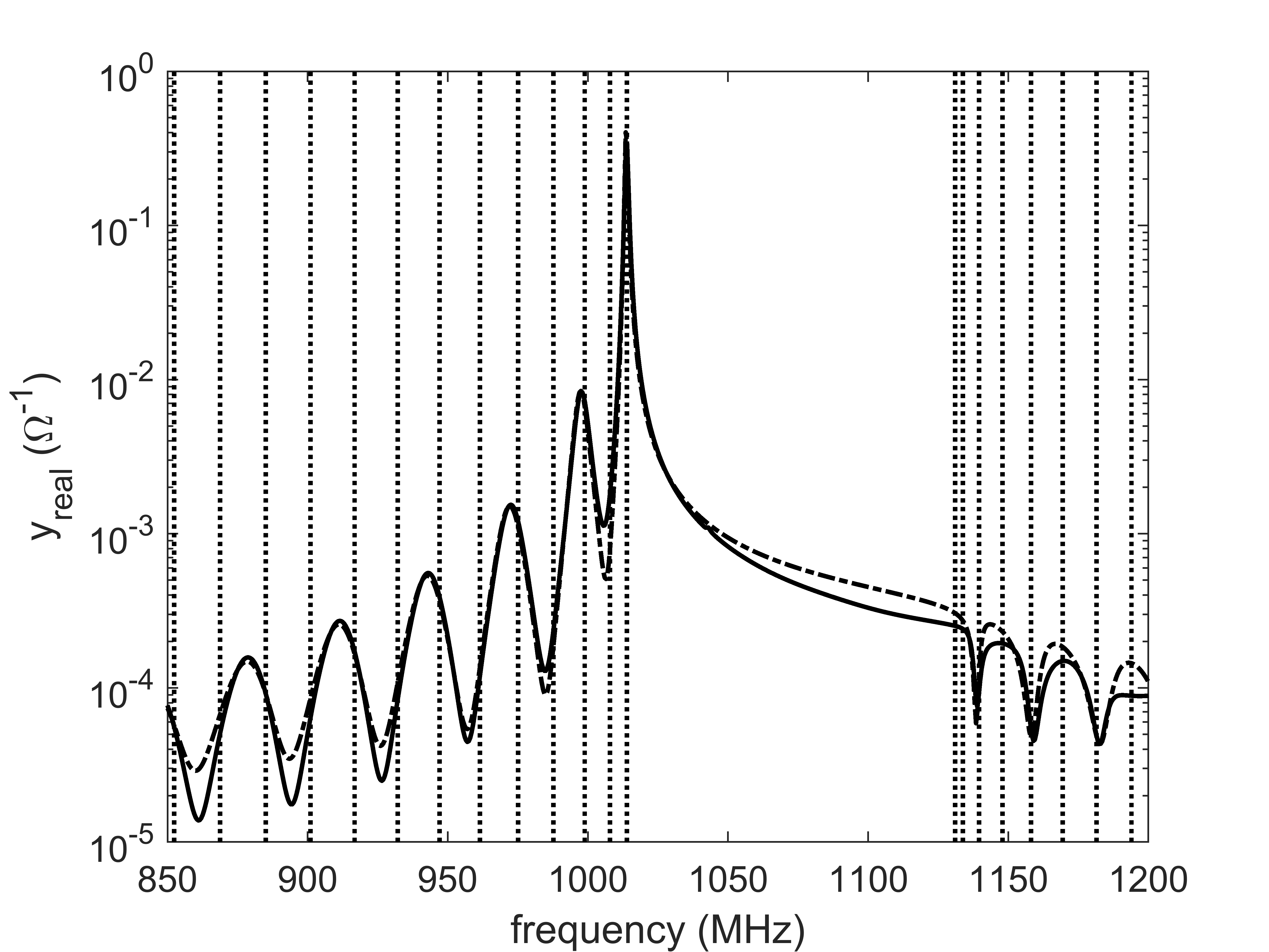}
	\end{subfigure}
	\begin{subfigure}[t]{0.5\textwidth}
		\includegraphics[trim=50 120 50 100,clip,width=1\textwidth]{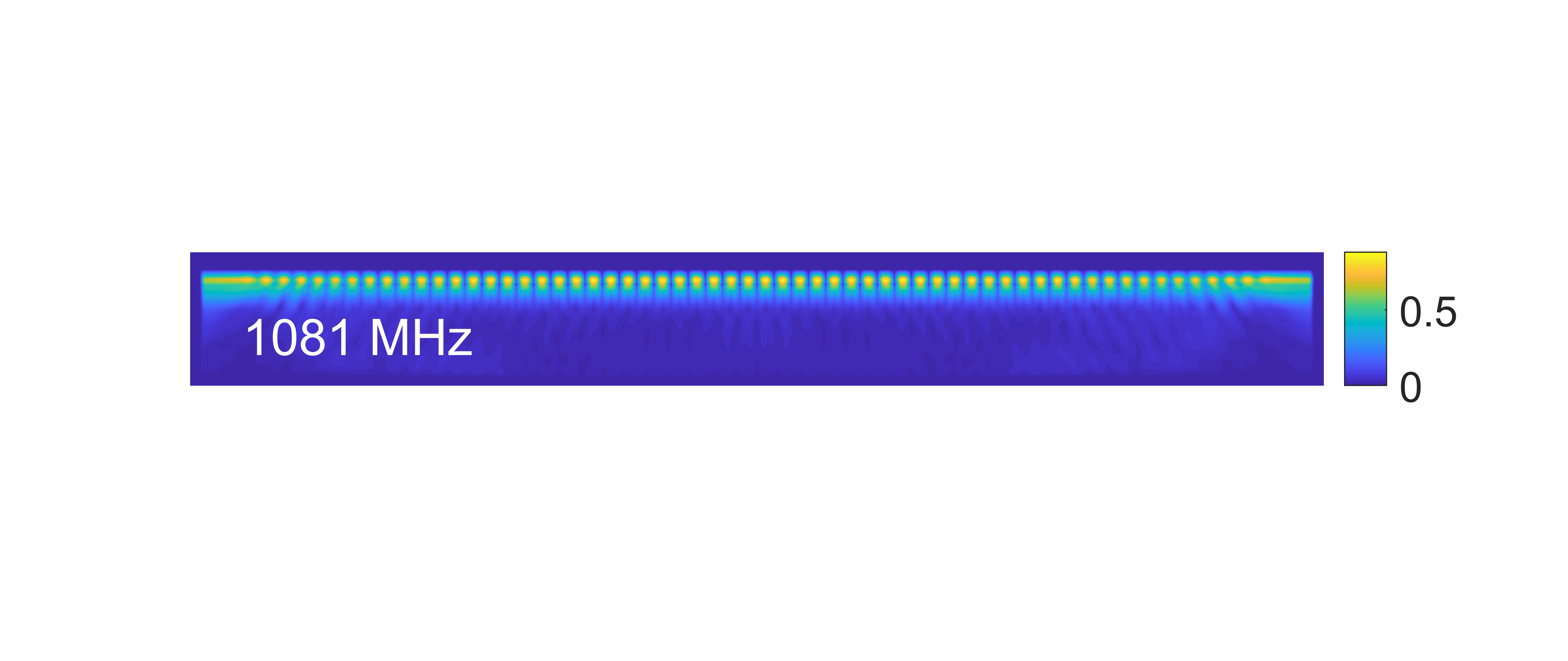}
	\end{subfigure}
\caption{(\textit{top}) Admittance of a synchronous IDT with 60 identical electrodes simulated with FEM (solid) and the lattice model simulation (dashed).  The peak in the magnitude of the admittance, $|Y|$ at 1015 MHz corresponds to the resonance and the minimum at 1055 MHz the anti-resonance. (\textit{middle}) The vertical dotted lines correspond to the mechanical eigenfrequencies calculated with the lattice model.  For this synchronous IDT, the peaks in the conductance occur near the even eigenfrequencies and reveal the acoustic band structure. The band gap is evident from 1015 to 1130 MHz. (\textit{bottom}) FEM simulation of the magnitude of the mechanical displacement (arb.units) for the full length of the IDT calculated at a frequency near the center of the band gap, 1081 MHz.  The displacements are confined to the surface and are evenly spread among the electrodes.}
\label{fig_y_synch}
\end{figure}

With a careful choice of parameters, the lattice model is very accurate.  The top plot of Figure \ref{fig_y_synch} shows $|y|$ of a synchronous IDT similar to those found in the microwave filters of mobile phones, simulated with the FEM analysis software(solid), Resonant ISN\textsuperscript{\textregistered} and with the lattice model (dashed).  The IDT is composed of 60 identical Cu electrodes 300 nm thick, deposited on 128$^{\circ}$YX-cut LiNbO$_3$ and covered by a thin layer of SiO$_2$, which is commonly done to reduce the sensitivity to temperature variations. The electrode widths are each 0.85 $\mu$m and are equally spaced by 0.85 $\mu$m.  The admittance of this IDT is essentially that of an approximately 1 pF capacitor, interrupted by a local maximum near 1015 MHz and minimum near 1055 MHz, which are the so-called resonance and anti-resonance, respectively.  

The middle plot of Fig. \ref{fig_y_synch} shows the conductance of both the FEM (solid) and lattice (dashed) simulations.  The vertical dotted lines indicate the mechanical eigenfrequencies calculated from Eq. 2.  The choice of electrical connection, $Y_{bus}$ determines which of the eigenmodes are coupled to the electrical system.  For a synchronous IDT, each peak in the conductance corresponds to an even mechanical eigenmode. The conductance is likely the most useful quantity to measure in simulations and experiments since each peak reveals an eigenmode, and collectively reveal the mechanical bandstructure.  Above the resonance frequency, the conductance is featureless until 1135 MHz, at which point the oscillations due to the cavity modes resumes.  The region from 1015 to 1135 MHz is the acoustic stopband or bandgap.  

The bottom plot of Fig. \ref{fig_y_synch} shows the FEM simulated magnitude of the mechanical displacement calculated at 1081 MHz, which is near the center of the band gap.  The displacements are confined to the surface and evenly spread along the entire length of the IDT.  

\begin{figure}[h!]
\centering
\includegraphics[width=0.5\textwidth]{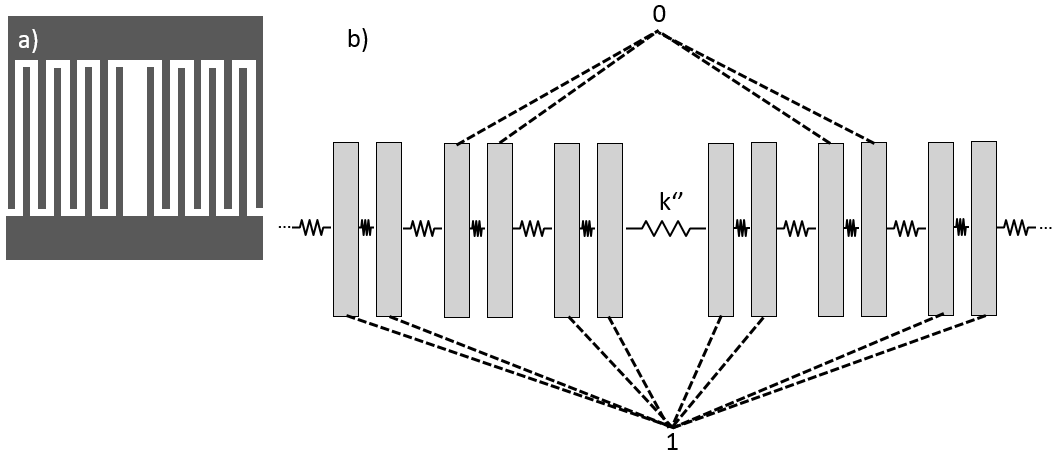}
\caption{(\textit{a}) Top-view schematic of hiccup IDT with every other electrode is electrically connected via a busbar, and one electrode removed near the center. (\textit{b}) Illustration of lattice model for hiccup resonator.  The setup is identical to Fig. \ref{schematic_synchronous}b, with the exception of $k''$, which represents the weaker coupling across the gap, i.e., $k''< k'$ and $k$.}
\label{schematic_hiccup}
\end{figure}

\begin{figure}[h!]
\centering
	\begin{subfigure}[t]{0.5\textwidth}
		\includegraphics[width=1\textwidth]{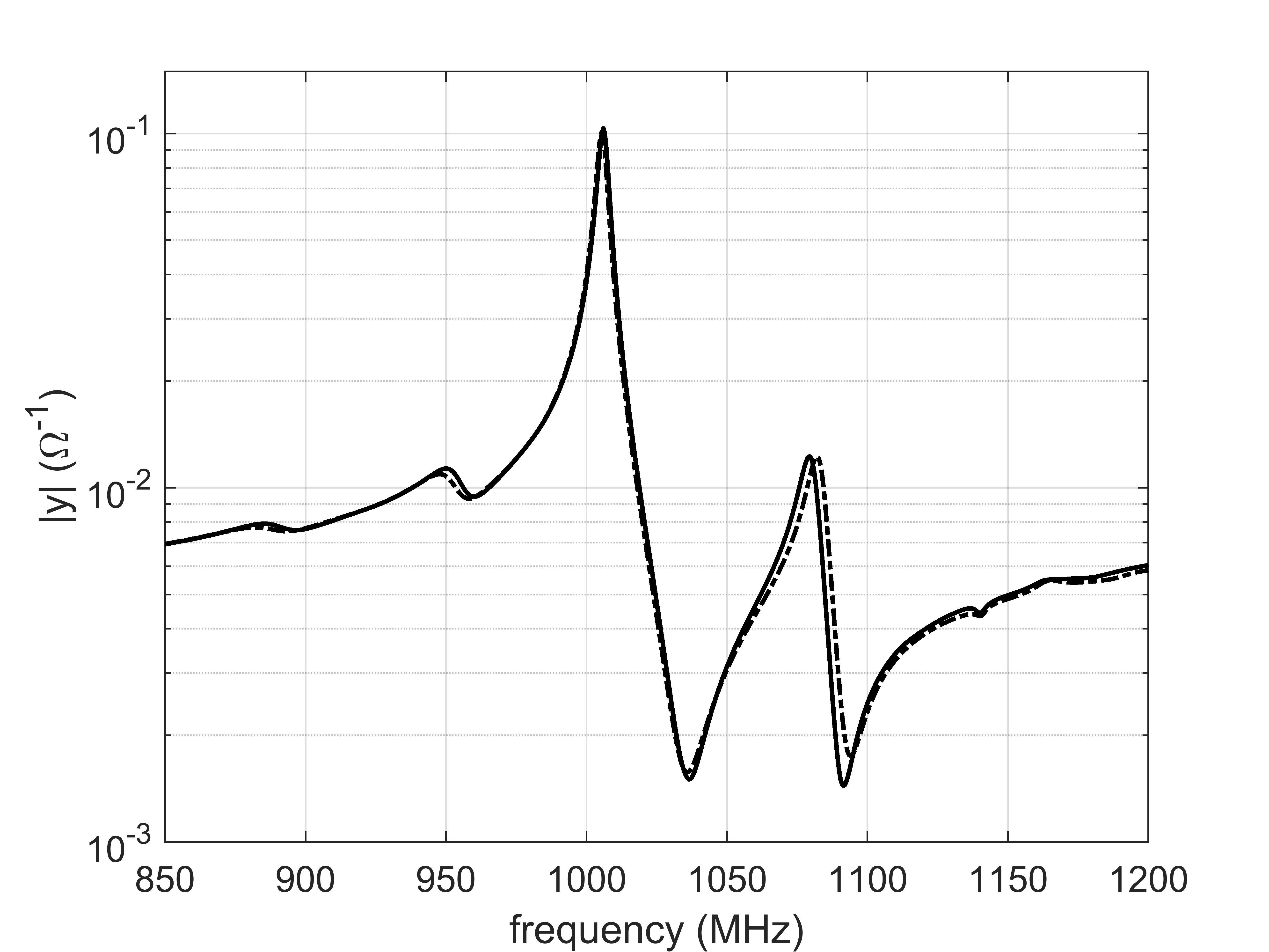}
	\end{subfigure}
	\begin{subfigure}[t]{0.5\textwidth}
		\includegraphics[width=1\textwidth]{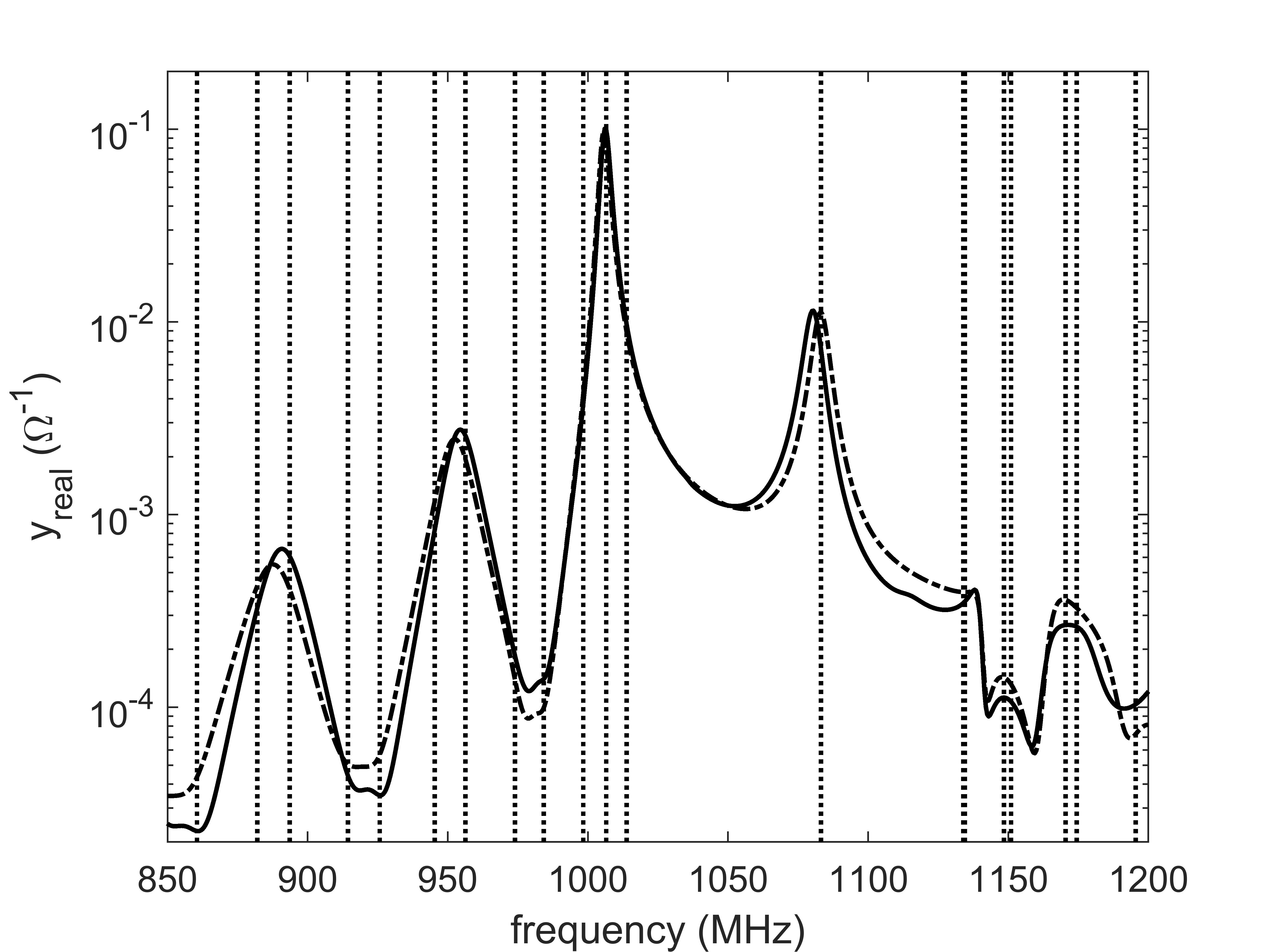}
	\end{subfigure}
	\begin{subfigure}[t]{0.5\textwidth}
		\includegraphics[trim=50 120 50 100,clip,width=1\textwidth]{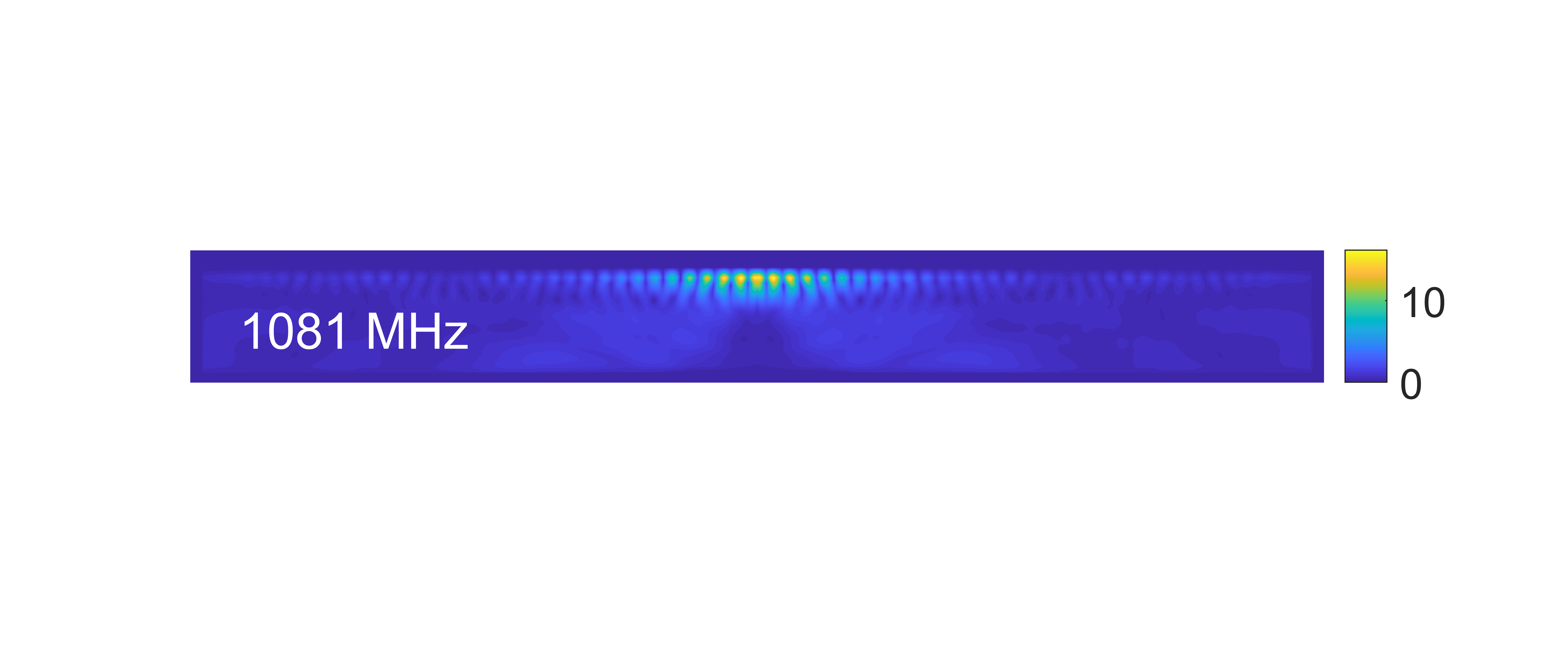}
	\end{subfigure}
\caption{(\textit{top}) Admittance of a hiccup resonator with 59 identical electrodes simulated with FEM (solid) and the lattice model simulation (dashed). Aside from the missing electrode near the center, the dimensions are the same as the IDT simulated in Fig. \ref{fig_y_synch}. The gap creates an additional hiccup resonance at 1081 MHz. (\textit{middle}) The vertical dotted lines show the mechanical eigenfrequencies calculated with the lattice model.  The hiccup resonance corresponds to an eigenmode near the center of the acoustic band gap. (\textit{bottom}) FEM simulation of the magnitude of the mechanical displacement (arb.units) for the full length of the IDT calculated at a frequency near the hiccup resonance at 1081 MHz.  The displacements are confined to the surface, but localized to the gap near the center of the IDT.}
\label{yu_hiccup}
\end{figure}

The change from a synchronous to a hiccup resonator is a small one; remove an electrode from the center.  Fig. \ref{schematic_hiccup}a illustrates a hiccup IDT.  The wide hiccup gap in the center can be modeled by introducing an addition weaker spring, $k''$ to the lattice model as shown in Fig. \ref{schematic_hiccup}.  Aside from this gap, the setup is identical to that of a synchronous IDT.  The top and middle plots of Figure \ref{yu_hiccup} show the results of a FEM (solid) and lattice (dashed) simulations for the hiccup resonator.  Again, the vertical dotted lines show the mechanical eigenfrequecies and a similar band gap to the synchronous resonator. The hiccup gap introduces an eigenfrequency and a peak in the conductance near the center of the band gap at 1081 MHz.  The mechancial displacement is calculated at 1081MHz from the FEM simulation.  It reveals that the mode is indeed localized at the hiccup gap as expected of an edge state. 

\begin{figure}[h!]
\centering
	\includegraphics[width=0.425\textwidth]{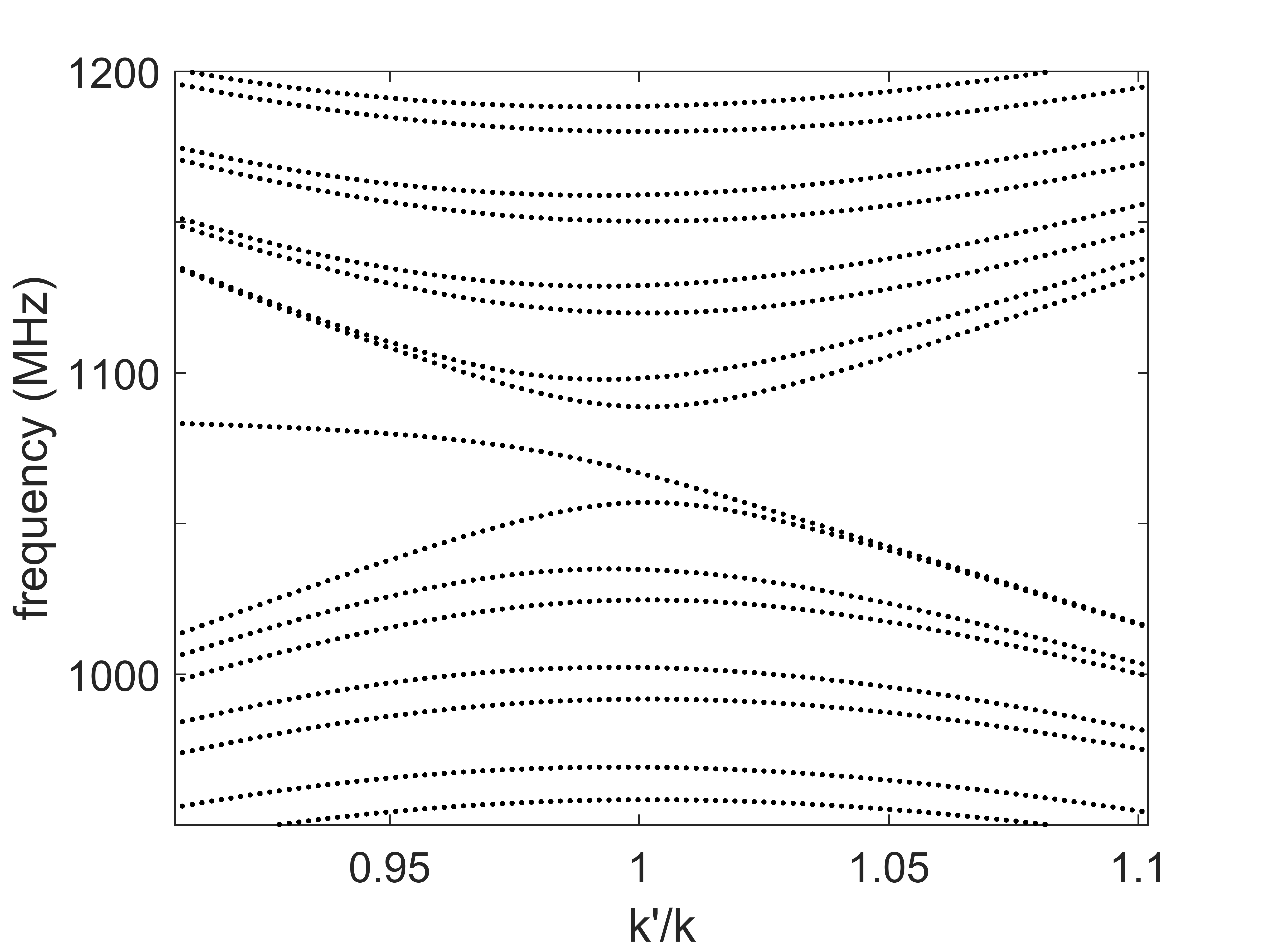}
\caption{Eigenfrequencies of a hiccup resonator calculated with the lattice model as a function of $k'/k$.  The eigenfrequencies at $k'/k = 0.9$ correspond to the model in Fig. \ref{yu_hiccup}, with the edge state at 1081 MHz.  As $k'/k$ increases to 1, the band gap decreases.  For $k'/k>1$, the band gap opens and forming a topologically distinct region without an edge state.}
\label{hiccup_bandgap}
\end{figure}

Localized edge modes can be also be created by prosaic translational symmetry breaking.  To prove that the hiccup resonance is a topologically protected edge state, the band gap must be tuned and examined for the presence of an eigenfrequency in the band gap.  Figure \ref{hiccup_bandgap} shows eigenfrequencies of the hiccup resonator as a function of $k'/k$.  For the hiccup resonator shown in Fig. \ref{yu_hiccup}, $k'/k = 0.9$.  There are two topologically distinct sectors. The sector without the edge state ($k'/k>1$), can only be reached by transitioning through a region where the band gap vanishes ($k'/k = 1$), which deomonstrates the topological protection of the edge state.  

\begin{figure}[h!]
\centering
	\begin{subfigure}[t]{0.5\textwidth}
		\includegraphics[width=1\textwidth]{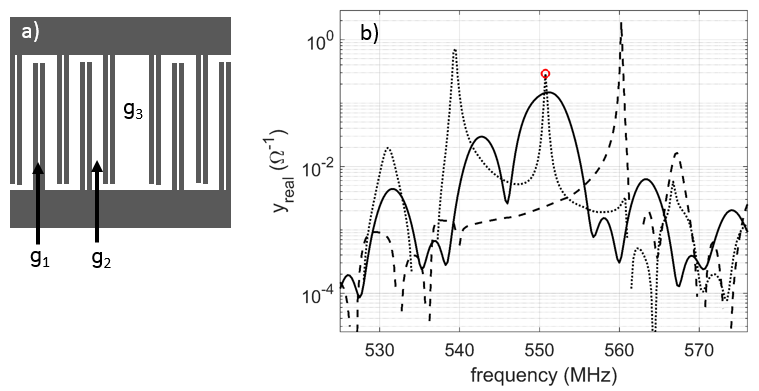}
	\end{subfigure}
	\begin{subfigure}[t]{0.5\textwidth}
		\includegraphics[trim=100 270 100 270,clip,width=1\textwidth]{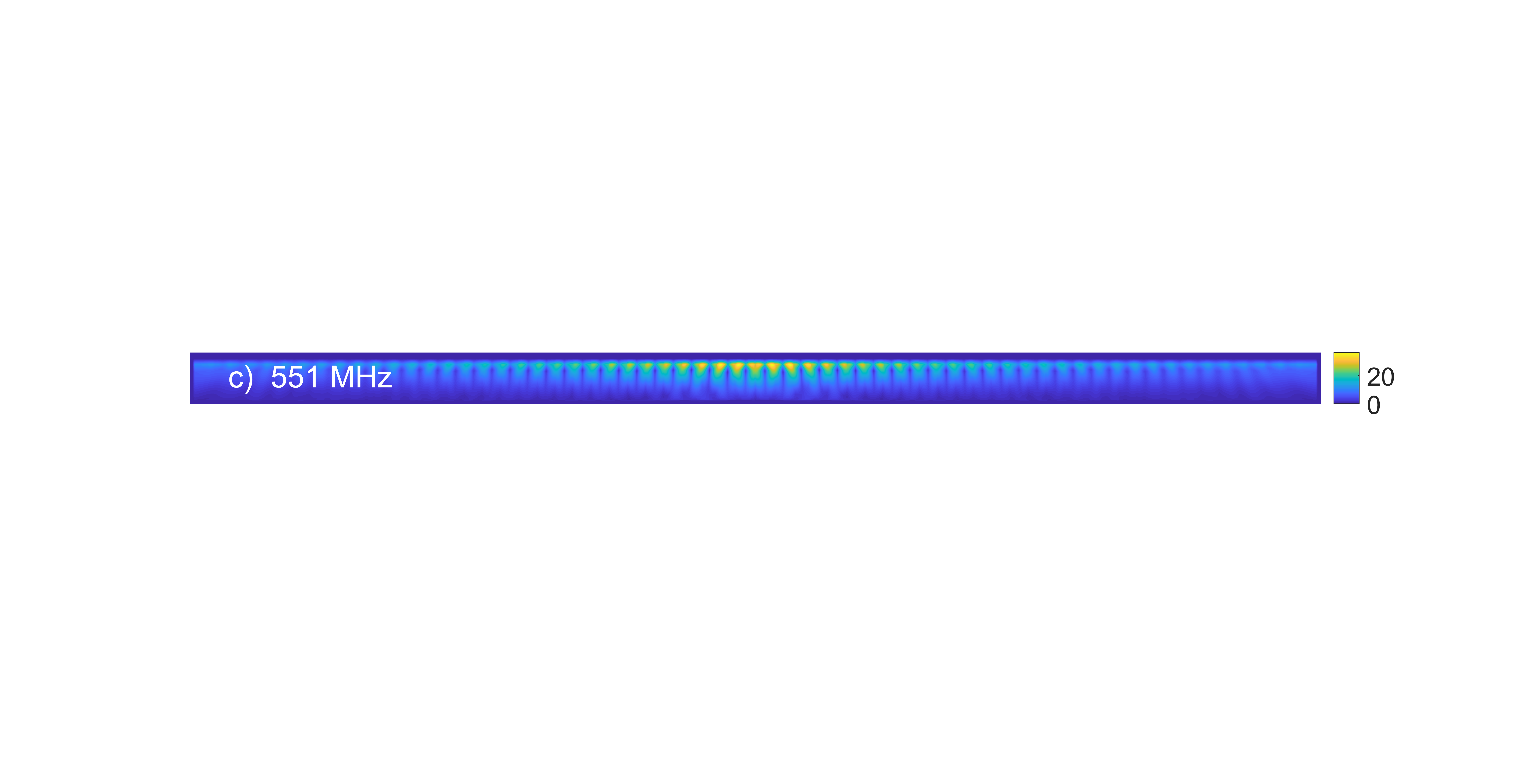}
	\end{subfigure}
	\begin{subfigure}[t]{0.5\textwidth}
		\includegraphics[trim=100 270 100 270,clip,width=1\textwidth]{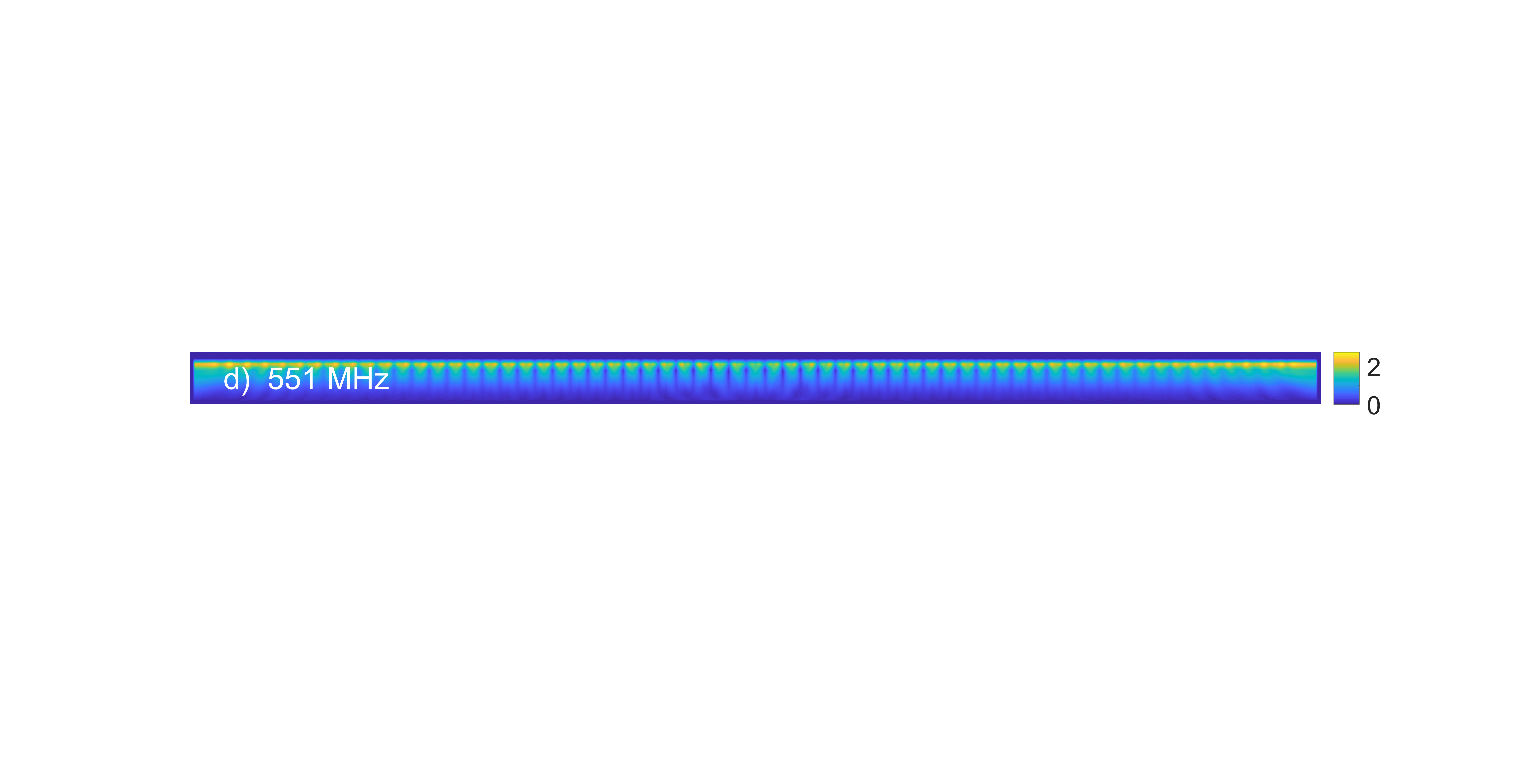}
	\end{subfigure}
	\begin{subfigure}[t]{0.5\textwidth}
		\includegraphics[width=0.85\textwidth]{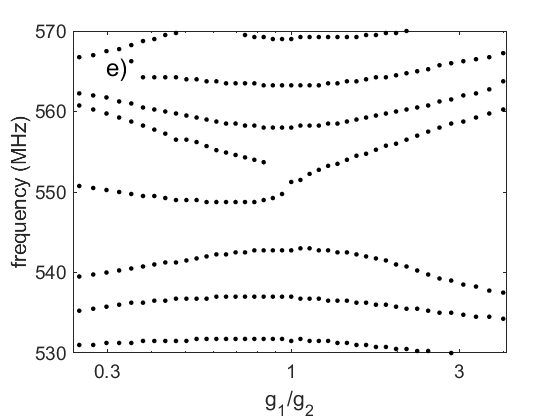}
	\end{subfigure}
\caption{(\textit{a}) Schematic of double-electrode hiccup resonator. The acoustic band gap can be tuned by varying the gaps, $g_1$ and $g_2$. (\textit{b}) FEM simulations were performed on an IDT with 398 electrodes, for values $1/4 \le g_1/g_2 \le 4$.  The conductance is shown for $g_1/g_2$ equal to 1/4 (dotted), 1 (solid), and 4 (dashed).  The red marker denotes the hiccup resonance, i.e., topological edge state, at 551 MHz near the middle of the band gap.  (\textit{c-d}) The mechanical displacements (arb.units) were calculated for $g_1/g_2 = 1/4$ (\textit{c}), which shows the mode localized near the hiccup gap.  For $g_1/g_2 = 4$ (\textit{d}), the displacements are small and uniform along the IDT.  (\textit{e}) Each marker corresponds to a peak in the conductance in a FEM simulation for $1/4 \le g_1/g_2 \le 4$.  The band gap closes and the topological edge state disappears as $g_1/g_2$ approaches 1.  For $g_1/g_2 \ge 1$, the band gap opens and there is no edges state.}
\label{schematic_doubleElectrode}
\end{figure}

Ideally, the physical parameters of the hiccup resonator could be varied to tune the band gap such that Fig. \ref{hiccup_bandgap} could be simulated with FEM.  However, the bandgap of a real synchronous or hiccup IDT cannot be varied over a sufficiently wide range for a clear demonstration.  As a proxy for the traditional hiccup geometry, it is therefore convenient to study the double electrode, hiccup IDT shown in \ref{schematic_doubleElectrode}a.  This can be thought of as a more literal realization of the lattice model with two electrodes per lattice site.  Keeping the electrode widths the same as the previous FEM simulations (0.85 $\mu$m), the gaps $g_1$ and $g_2$ can be varied over a sufficiently large range and realize the same features as the conventional hiccup IDT.  

Fig. \ref{schematic_doubleElectrode}(b-e) shows the results of FEM simulations for this proxy resonator with 398 electrodes.  The number was increased from 59 due to the significantly greater damping when compared to a conventional single electrode resonator.  Fig. \ref{schematic_doubleElectrode}b shows the conductance for $g_1/g_2$ = 1/4 (dotted), 1 (solid), and 4 (dashed) all with the same value of $g_3=4.25\mu$m. The peaks in the conductance reveal the mechanical band structure.  The dotted line shows a wide band gap with the hiccup resonance at 551 MHz denoted with the red marker.  There is no apparent band gap for the solid line.  And for the dashed line, the band gap is again wide, but without a mode at the center of the band gap.  Fig. \ref{schematic_doubleElectrode}c confirms that the the hiccup mode is localized to the region near the domain wall for $g_1/g_2=1/4$.  (Note, since the number of electrodes is so large, only the center half is shown.)   The mechanical displacements are small and uniform at 551 MHz for $g_1/g_2 = 4$, as shown in Fig. \ref{schematic_doubleElectrode}d.  

Finally, the peaks in the conductance were tabulated from FEM simulations for 1/4$\le g_1/g_2 \le 4$.  Fig. \ref{schematic_doubleElectrode}e shows the frequencies for these peaks.  As in Fig. \ref{hiccup_bandgap}, and as is expected for the 1D SSH model, there are topologically distinct regions with the topological edge state existing only for $g_1/g_2 \le 1$. 

Despite the simplicity of the lattice model, it fares well when compared to highly accurate FEM simulations.  In addition, the lattice model facilitates the mapping of the simplest topological mechanical metamaterials onto the physics of an IDT.  Considering a significant motivation for studying topological mechanical metamaterials is the desire for novel technological applications, it should be satisfying to see in retrospect they have in fact been so useful for the past three decades.

% If you have acknowledgments, this puts in the proper section head.
\begin{acknowledgments}
\section{Acknowledgment}
I thank Greg Dyer and Patrick Turner for useful discussions and suggestions.
\end{acknowledgments}

%%%%%%%%%%%%%%%%%%%%%%%%%%%%%%%%%%%%%%%%%%%%%%%%%%%%%%%%%%%%%%%
%%%%%%%%%%%%%%%%%%%%%%%%%%%%%%%%%%%%%%%%%%%%%%%%%%%%%%%%%%%%%%%
%%%%%%%%%%%%%%%%%%%%%%%%%%%%%%%%%%%%%%%%%%%%%%%%%%%%%%%%%%%%%%%
% Specify following sections are appendices. Use \appendix* if there
% only one appendix.
\appendix*
\section{Appendix}
\begin{figure}[h!]
\centering
\includegraphics[width=0.4\textwidth]{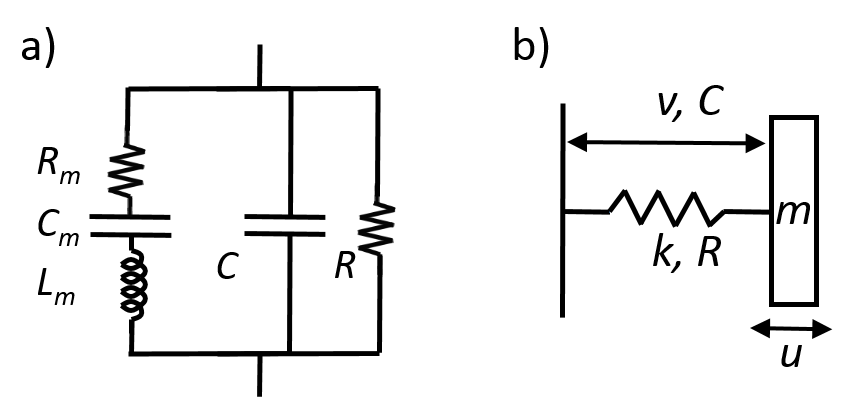}
\caption{(\textit{a}) The BvD model equivalent circuit, with one degree of freedom (either the voltage $v$ or current $I$) that captures the gross electrical features of an IDT. (\textit{b}) A simple electro-mechanical model with an additional degree of freedom, $u$.  A harmonic oscillator with mass ($m$), spring ($k$), and damping ($\tau$) forms a capacitor ($C$) in parallel with a resistor ($R$).  The two degrees of freedom are coupled with $\alpha$, the electro-mechanical coupling constant.}
\label{preBvD_circuit}
\end{figure}

The equivalent circuit shown in \ref{preBvD_circuit}a, known as the modified Butterworth-van Dyke (mBvD) model \cite{Butterworth, vanDyke}, captures the essential electrical features of an IDT; wide-band capacitor with a narrow band resonance at $\omega = \frac{1}{\sqrt{L_mC_m}}$ and antiresonance at $\omega = \frac{1}{\sqrt{L_mC_m + L_mC}}$.  Them BvD model can be derived from the simple electro-mechanical model depicted in \ref{preBvD_circuit}b \cite{McHughPRA}.  A single mechanical degree of freedom is introduced, $u$, which is the displacement from equilibrium. The electrical resistance is assumed to be large, i.e., $i\omega C \gg 1/R$.  Assume also that $|u|$ is small enough such that the capacitance may be considered constant. The voltage across $C$ is then proportional to the charge and due to the piezoelectricity,  the mechanical displacement, i.e., $Cv = Q + \alpha u.$ $\alpha$ is the electromechanical coupling constant and has dimensions $\bf{TIL^{-1}}$.  Since it is more convenient to work with currents, $I$, rather than charge, we take the time derivative of both sides giving
\[
C\frac{dv}{dt} = I + \alpha \frac{du}{dt}.
\]

The mechanical degree of freedom may be treated as a mass, $m$, on a spring with constant $k$ forming a damped harmonic oscillator with damping $\tau$. Along with any explicit mechanical driving force, $J$, there is a driving force proportional to $v$, which gives the equation of motion for $u$,
\[
m\frac{d^2u}{dt^2}=-ku +J - \alpha v -\tau\frac{du}{dt}.
\]
One of the primary virtues of using piezoelectric IDTs is they allow the acoustics to be sensed and excited electrically.  Therefore, all external mechanical driving forces are assumed to be zero, i.e., $J=0$.

It is convenient to write these equations of motion for $v$ and $u$ in the frequency domain
\begin{eqnarray}
\label{aeq1}
i\omega Cv = I + i\omega\alpha u\\
\label{aeq2}
(k+i\omega\tau-m\omega^2)u= - \alpha v.
\end{eqnarray}
These can be solved to give an expression for v alone,
\begin{equation}
\label{aeq3}
\left(i\omega C + \frac{i\omega\alpha^2}{k+i\omega\tau-m\omega^2}\right)v = I .
\end{equation}
This has the form of Ohm's law and the electrical admittance is equivalent to the mBvD model provided $L_m = m/\alpha^2$, $C_m = \alpha^2/k$, and $R_m = m\tau/\alpha^2$.  The electrical resonance frequency is set by the mechanical resonance, $\omega_0=\sqrt{k/m}$.

Note, for electrical circuit analysis either the voltage or current can serve as the driving force.  Accordingly, the electrical admittance is independent of this choice.  However, the choice does affect the mechanical displacement.  If $v$ is the driving force, and $I$ is the electrical degree of freedom, the frequency dependence of $u$, which is given by Eq. \ref{aeq2}, clearly has a resonance at $\omega_0$. Conversely, if $I$ is the driving force, and $v$ is the electrical degree of freedom, there is an mechanical resonance at $\omega = \sqrt{\frac{k}{m}+\frac{\alpha^2}{mC}}$.  To see this, solve Eq. \ref{aeq3} for $v$ and substite into Eq. \ref{aeq2} to give, 
\begin{equation}
u = \frac{-\frac{\alpha}{i\omega C}I}{k+i\omega\tau-m\omega^2+\alpha^2/C}.
\end{equation}
In real microwave measurements, the source and matching impedences will result in behavior between these two ideal cases.  For the work above, $v$ iss the electrical driving force, and $I$ is the degree of freedom.   

The basis of the lattice model is to apply these ideas to each electrode of an IDT.  Consider the array of $N$ electrodes shown in Fig. \ref{schematic_synchronous}a.  The displacements and voltages are defined with respect to the neighboring electrodes giving the equations of motion for each
\begin{widetext}
\begin{eqnarray}
i\omega C_n\left( v_n-v_{n+1}\right) + i\omega C_{n-1}\left( v_n-v_{n-1}\right) = I_n + i\omega\alpha_n \left( u_n - u_{n+1}\right)+ i\omega\alpha_{n-1} \left( u_n - u_{n-1}\right)\\
k_n(u_n-u_{n+1})+k_{n-1}(u_n-u_{n-1}) +i\omega\tau_n u_n-m_n\omega^2u_n = - \alpha_n( v_n-v_{n+1}) - \alpha_{n-1}( v_n-v_{n-1}).
\end{eqnarray}
\end{widetext}
Note, the acoustic, electric, and piezoelectric coupling to the left and right are not assumed to be the same, e.g., we distinguish $k_n$ from $k_{n-1}$.  Also, only nearest-neighbor interactions are considered.  The above system of coupled equations are written more compactly as,
\begin{eqnarray}
i\omega C\mathbf{v} =\mathbf{I} + i\omega\mathbf{\alpha u} \\
(k+i\omega\tau-m\omega^2)\bf {u}= - \alpha \mathbf{v},
\end{eqnarray}
where $\bf{u}$ = $(u_0, u_1,...u_{N-1})$ and $\bf{I}$ = $(I_0, I_1,...I_{N-1})$ are the degrees of freedom. $\bf{v}$ = $(v_0, v_1,...v_{N-1})$ are the driving voltages. $C, \alpha, k, \tau$, and $m$ are now considered $N\times N$ matrices.

Electrical connections must be added in addition to the capacitive coupling between electrodes.  Fig. \ref{schematic_synchronous}b shows how the electrodes may be connected to simulate the electrical coupling to the busbar. Every other finger connected to the common electrical nodes, $0$ and $1$, by an admittance $y_b$, which can be treated as a simple conductor.  Like the capacitance matrix $C$, an admittance matrix $Y_{bus}$ may be formed representing this additional electrical coupling between the electrodes and nodes.  Including $Y_{bus}$ gives
\begin{eqnarray}
\label{aeq4}
i\omega C\mathbf{v} + Y_{bus}\mathbf{v}=\mathbf{I} + i\omega\mathbf{\alpha u}\\
\label{aeq5}
\left( k+i\omega\tau-m\omega^2\right)\mathbf {u}= -\alpha \mathbf{v}.
\end{eqnarray}
$C$ could rightly be combined with $Y_{bus}$ to form a total admittance matrix, but it is kept separate here to emphasize the physical origin. Eqs. \ref{aeq4} and \ref{aeq5} can be solved for $\bf{I}$ to give
\begin{equation}
\mathbf{I} = Y\mathbf{v},
\end{equation}
where $Y =  i\omega C + Y_{bus}+i\omega\alpha Y_{mech}^{-1}\alpha$ is the total electrical admittance.

In this state, the lattice model has only one mechanical degree of freedom per lattice site, and will have no band gap.  For the simulations of the synchronous and hiccup resonators above, each electrode is treated as two identical masses coupled by spring, $k$.  If $k\neq'$ a band gap is formed. This can be realized more literally with the double electrode configuration of Fig. \ref{schematic_doubleElectrode}a. Incidentally, additional electrodes could be included per lattice site, which which allows for variations on the SSH model\cite{Maffei}.

To improve the accuracy of the lattice model over a wide frequency range, it is necessary to make two additional ad hoc modifications. A small contribution from the next-nearest neighbor interactions must be added to $k$.  For the synchronous and hiccup IDTs simulated above, it was necessary to add $k_{\mbox{nnn}}\approx -0.05k$.  Modifying the structure to enhance this long-range coupling may lead to interesting phenomena\cite{An, PerezGonzalez}. The second modification necessary for the lattice model is increase the damping of the electrodes near the ends of the IDT.  This necessary due to the escaping acoustic waves.

\bibliography{hiccup}

\end{document}